\documentclass[sigconf]{acmart}

\newcommand{\paperTitle}{Rethinking Similarity Search: Embracing Smarter Mechanisms over Smarter Data}

\newcommand{\sys}{\mbox{\it{VidQL}}\xspace}
\newcommand{\multi}{\mbox{multi-body}\xspace}
\newcommand{\Multi}{\mbox{Multi-body}\xspace}

\usepackage{anyfontsize}
\usepackage[hyphens]{url}

\definecolor{linkcolor}{HTML}{647382}
\definecolor{citecolor}{HTML}{647382} %
\definecolor{urlcolor}{rgb}{0.4,0.2,0.2}
\definecolor{sqlcolor}{HTML}{965d67}
\definecolor{smtcolor}{HTML}{5d968c}
\definecolor{webblue}{rgb}{0,0,.7}
\definecolor{webgreen}{rgb}{0,.5,0}
\definecolor{webbrown}{rgb}{.6,0,0}
\definecolor{notecolor}{HTML}{FFF8DC}

\usepackage{amsmath,amsopn,amssymb}
\usepackage[cal=boondoxo]{mathalfa}
\usepackage{subcaption}
\usepackage{endnotes,xspace,graphicx,fancyvrb,multirow}
\usepackage{supertabular,booktabs}
\usepackage{array,underscore, relsize}
\usepackage{fancyhdr}
\usepackage{enumitem}
\usepackage{booktabs}
\usepackage{pifont}
\usepackage{listings}
\usepackage{paralist}

\usepackage{multirow}
\usepackage[scaled]{beramono}
\usepackage{tabularx}
\usepackage{multirow}

\makeatletter
\newcommand\BeraMonottfamily{%
  \def\fvm@Scale{0.85}%
  \fontfamily{fvm}\selectfont%
}
\makeatother

\definecolor{mymauve}{rgb}{0.58,0,0.82}

\lstdefinestyle{SQLStyle}{
  language=SQL,
  basicstyle={\small\ttfamily},
  breaklines=true,
  frame=none,
  numbers=none,
  keepspaces=true,
  captionpos=b,
  stringstyle=\color{mymauve},
  keywordstyle=\color{blue},
  commentstyle=\color{dkgreen},
}

\lstdefinestyle{ScriStyle}{
language=SQL,
basicstyle=\BeraMonottfamily\footnotesize, 
keywordstyle=\color{smtcolor}\bfseries,
morekeywords={and, or, not},
aboveskip = 0.05in,
belowskip = 0.05in,
literate = {-}{-}1, %
}

\usepackage{aliascnt}

\usepackage[ruled,linesnumbered,vlined]{algorithm2e}
\usepackage{setspace}
\usepackage{enumitem}
\usepackage[color=notecolor]{todonotes} %

\SetAlFnt{\small}
\SetAlCapFnt{\small}
\SetAlCapNameFnt{\small}

\usepackage{semantic}

\usepackage{stmaryrd}
\usepackage{ltablex}
\usepackage{mathtools}
\usepackage{adjustbox}
\usepackage[group-separator={,}]{siunitx}
\usepackage[capitalize,noabbrev,nameinlink]{cleveref}

\crefname{lstlisting}{listing}{listings}
\Crefname{lstlisting}{Listing}{Listings}

\usepackage{chngcntr}
\counterwithout{equation}{section}

\newcommand{\hide}[1]{}

\newcommand{\PPP}[1]{
\vspace{0.05in}
\noindent{\textit{\IfEndWith{#1}{.}{#1}{#1.}}}
}

\newcommand{\squishitemize}{
 \begin{list}{$\bullet$}
  { \setlength{\itemsep}{0pt}
     \setlength{\parsep}{0pt}
     \setlength{\topsep}{0pt}
     \setlength{\partopsep}{0pt}
     \setlength{\leftmargin}{1.95em}
     \setlength{\labelwidth}{1.5em}
     \setlength{\labelsep}{0.5em} } }

\newcounter{Lcount}
\newcommand{\squishlist}{
    \begin{list}{\arabic{Lcount}. }
   { \usecounter{Lcount}
        \setlength{\itemsep}{0pt}
        \setlength{\parsep}{3pt}
        \setlength{\topsep}{0pt}
        \setlength{\partopsep}{0pt}
        \setlength{\leftmargin}{2em}
        \setlength{\labelwidth}{1.5em}
        \setlength{\labelsep}{0.5em} } }

\newcommand{\squishend}{\end{list}}

\newcommand{\bit}{\begin{compactitem}}
\newcommand{\eit}{\end{compactitem}}
\newcommand{\ben}{\begin{compactenum}}
\newcommand{\een}{\end{compactenum}}

\definecolor{dkgreen}{rgb}{0,0.6,0}

\def\Snospace~{\S{}}

\newcommand{\minihead}[1]{{\vspace{.45em}\noindent\textbf{#1.} }}

\title{\paperTitle}

\begin{document}

\author{Renzhi Wu, Jingfan Meng, Jie Jeff Xu, Huayi Wang, Kexin Rong}
\affiliation{Georgia Institute of Technology}

\begin{abstract}

In this vision paper, we propose a shift in perspective for improving the effectiveness of similarity search. Rather than focusing solely on enhancing the data quality, particularly machine learning-generated embeddings, we advocate for a more comprehensive approach that also enhances the underpinning search mechanisms. We highlight three novel avenues that call for a redefinition of the similarity search problem: exploiting implicit data structures and distributions, engaging users in an iterative feedback loop, and moving beyond a single query vector. These novel pathways have gained relevance in emerging applications such as large-scale language models, video clip retrieval, and data labeling. We discuss the corresponding research challenges posed by these new problem areas and share insights from our preliminary discoveries.

\end{abstract}

\maketitle

\section{Introduction}
\label{sec:introduction}

Similarity search studies the problem of finding the most pertinent data points in a database when compared to a specific query point. A na\"ive approach to this problem involves exhaustive search through the database for each query, which poses computational challenges for large datasets. To overcome these challenges, researchers have developed efficient indexing techniques, including methods like locality sensitive hashing~\cite{datar2004locality,lv2007multi}, nearest neighbor graphs~\cite{dong2011efficient,malkov2018efficient}, and product quantization~\cite{jegou2010product,ge2013optimized}, that improve computational efficiency by intelligently reducing the search space to a small subset of the dataset. These techniques have become vital to the practical application of similarity search in large datasets. 

Further, machine learning (ML) algorithms have greatly expanded the scope of similarity search. Modern ML models can transform various forms of structured and unstructured data, such as text, image, and time series, into embeddings. These high dimensional vectors are trained such that more similar inputs cluster closer together. This transformation allows users to search for abstract or semantic entities, such as searching for visually similar artworks, using distances in the vector embedding space as a proxy for similarity. By quantifying these formerly abstract concepts, ML has opened up a plethora of new applications for similarity search.

Despite these leaps forward, we have observed a surprising stagnation in the development of the similarity search problem formulation itself. Much of the progress seems to be driven by the machine learning community's endeavors to create superior machine models that offer optimal data representation or embeddings. While vector databases have begun to explore advance features such as metadata filtering, which adds constraints on non-vector data to the vector search, the focus of most current research remains on the ``basic'' similarity search problem. Our view is that improving data quality should not be the only method to enhance the similarity search effectiveness; an equally promising approach is to leverage smarter search mechanisms. We believe that the database community has much to contribute by exploring advanced query processing techniques beyond the basic similarity search problem. 

In this paper, we highlight several opportunities that could serve as potential avenues for future research.

\minihead{Opportunity 1: Leveraging Implicit Structures in Data} Embeddings are not arbitrary high-dimensional vectors; they are trained to encapsulate semantic meanings. Therefore, these vectors often contain implicit structure. For instance, embeddings associated with artworks may encapsulate diverse artistic styles, so that artworks of similar styles cluster together in the high-dimensional vector space. While these structures may not be explicitly encoded in metadata, leveraging them during similarity searches could help enhance the search quality. For instance, the search could be adjusted to ensure that the nearest neighbor is not only close to the query in terms of distance but also belongs to the same class or cluster. The question remains in how we can effectively incorporate such implicit structure into the search objectives. 

\minihead{Opportunity 2: Engaging Users in the Loop}
User feedback is a critical resource for improving the accuracy of similarity search results. While it is possible to retrain the machine learning model to improve the quality of embeddings based on additional labels provided by the user, doing so incurs significant computation overhead. Users may also have domain-specific knowledge, such as in the form of user-defined functions, which could improve the efficiency and accuracy of the similarity search. However, this knowledge is not currently captured in the pre-trained embedding space. Ideally, we would like to dynamically adapt the similarity search results according to user feedback for a specific query or question domain without updating the entire database. 

\minihead{Opportunity 3: Searching under Multi-object Constraints} Traditional similarity searches focus on identifying data points similar to a single query point. Increasingly, applications require measuring similarity between {\em sets} of objects rather than individual ones. Simply retrieving the top-$k$ match for each object can lead to low recall for applications that contain complex semantic and spatial relationships between objects. For example, we might want to retrieve images with visual concepts of the people and traffic lights where the person is standing below the traffic light. We need to revisit indexing and search algorithms to effectively handle these inter-object constraints.

These opportunities arise from emerging applications such as conversational chatbots, video clip retrieval, and data labeling. In the subsequent sections of this paper, we delve into the specific applications that give rise to these novel problem formulations in similarity search. We also discuss the research challenges associated with these new problems and share our initial findings.

\section{Background}

\label{sec:advance}
Vector databases are designed to store, manage, and perform similarity searches on high-dimensional vectors. Apart from the basic similarity search, these databases have begun to develop more advanced search features. We discuss two such features below.

\minihead{Metadata filtering} Similarity search queries on vector data are often used along with additional filters on non-vector metadata~\cite{metafilter1, metafilter2}. For example, customers of an e-commerce platform might want to search for items that are visually similar to a given image (vector data) and contain specific keywords in the product name (non-vector data). There are different plans for executing these hybrid queries: for example, the system can first filter based on metadata and then retrieve top-$k$ similarity search results, or it can first retrieve the top-$\alpha \cdot k (\alpha > 1)$ results and then apply the metadata filters. The optimal physical plan vary depending on selectivity of the vector and non-vector conditions~\cite{wei2020analyticdb}.

\minihead{Multi-vector Query} In certain applications, the similarity is computed based on not one but a set of vectors. Consider a video surveillance system that represents each person $X$ captured on camera using several vectors: $v_0$ represents front face features, $v_1$ represents side face and $v_2$ represents posture. The similarity between two people is computed by the weighted sum ($g$) of the inner product ($f$) between each pair of corresponding feature vectors: $g(f(X.v_0, Y.v_0), f(X.v_1, Y.v_1), f(X.v_2, Y.v_2))$. A straightforward approach to support this query is to independently retrieve the top-$k$ results for each feature vector, but this can lead to low recall. Prior work has begun to explore alternative strategies like merging multiple vectors into one for decomposable similarity functions $f$, or iteratively conducting a top-$k'$ query and incrementing $k'$~\cite{milvus}. Nevertheless, efficiently and accurately handling multi-vector queries remains an open research question.

\section{Distribution-aware Search}
In this section, we explore the potential of using the inherent structure within vector data to enhance the quality of similarity searches. This involves leveraging both the local distribution surrounding the query (\S~\ref{sec:local}) and the global distribution of the dataset (\S~\ref{sec:global}).

\subsection{Application Scenarios}

\minihead{Time-series Subsequence Retrieval} A common task in time series applications is to identify and retrieve subsequences that are from the same class as the given query subsequences. Consider a dataset containing users' desktop activities logged as time-stamped mouse click events. Suppose we have a sequence of these click events associated with the task of sending an email and we want to retrieve other subsequences that represent the same activity.

The typical similarity-based approach would only retrieve subsequences with the smallest distances to the query. However, this approach does not ensure that the retrieved subsequences belong to the same class and can be enhanced by leveraging local data distribution. We expect subsequences from the same class to likely locate in the same cluster. Therefore, a more effective method would not only retrieve subsequences that have a small distance to the query but also belong to the same local cluster.

\minihead{Custom Q\&A System} A generic, pre-trained Large Language Model (LLM) can be adapted to reflect an individual organization's specific internal knowledge via fine-tuning. A more cost-effective approach is to utilize vector databases as external memory for retrieving relevant context and improving LLM prompting.

We describe a typical workflow for users interested in building a Q\&A system on custom datasets. During the pre-processing phase, all proprietary documents are indexed by generating an embedding for each one and storing them in a vector database. During the querying phase, we generate embedding for the user query and perform a similarity search on the vector database to retrieve the top-$k$ most relevant documents with respect to the query. These documents, alongside the original query, are then used as context in the LLM to produce a response.

Empirical findings from industry practitioners show that leveraging the global data distribution can help improve retrieval performance~\cite{knnvssvm, Understa73, svm-langchain}, which in turn improves the quality of the custom Q\&A system. Essentially, by examining the query's position within the global distribution, we can pinpoint its unique characteristics within the dataset. This allows us to retrieve context that is both more relevant and personalized to the query. \vspace{-1em}

\subsection{Leveraging Local Distribution}
\label{sec:local}

We propose to incorporate constraints on local clustering structures into the objective function of similarity search. The idea is to perform the similarity search in an iterative manner, gradually expanding the query set by adding the newly identified nearest neighbor at each iteration. As such, it requires each nearest neighbor to be close not only to the original query vector but also to all previously identified nearest neighbors. This encourages the retrieval of nearest neighbors that are tightly clustered.

\begin{figure}[t]
    \centering
\includegraphics[width=\linewidth]{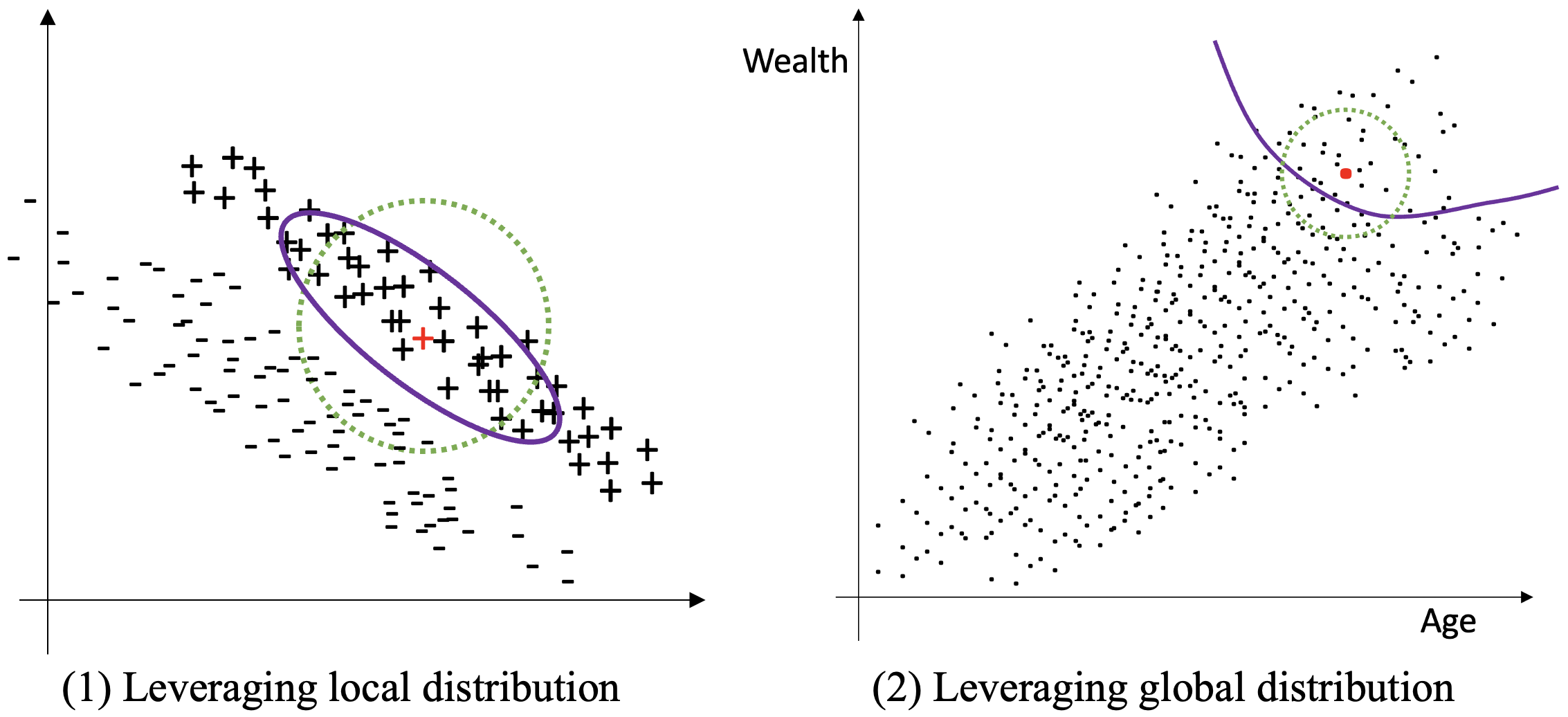}
\caption{Traditional similarity search retrieves points in green dotted circle in both figures. (1) The red plus \textcolor{red}{+} is the query. Our proposed approach retrieves points in the purple solid circle. (2) Leveraging global distribution. The red dot \textcolor{red}{$\cdot$} is the query. An SVM-based approach retrieves points on the top right of the purple solid curve. }
\label{fig:distri}
\end{figure}

Figure~\ref{fig:distri} (1) illustrates the effect of our proposed approach. In this example, data points form two skewed clusters, indicated by the $+$ and $-$ signs. When we conduct a similarity search for the query (\textcolor{red}{+} in the figure), the traditional approach (green dotted circle) retrieves many data points from the opposing cluster, while our proposed method (purple solid circle) primarily retrieves data points from the same cluster as the query.

The search procedure proceeds as follows:
\squishitemize
 \item For each data item in the dataset, calculate the sum of distances to all queries in the current query set. Select the data item with the smallest sum as the nearest neighbor.
    \item Update the query set with the newly identified near neighbor.
    \item Repeat until a predefined stopping criterion is met.
\squishend
   
\minihead{Preliminary Results}
We evaluated our approach on a private desktop activity dataset comprising interaction logs (e.g., mouse clicks) from 5 users over two days. Given one template log sequence representing a certain task, e.g., sending an email, we aim to retrieve subsequences of the same task. There are 3 tasks of interest with 72 instances of the tasks in total. 
The similarity function between two sequences $S_1$ and $S_2$ (i.e. $f_{\text{sim}}(S_1, S_2)$) is provided by a domain expert.
We consider sliding window similarity search with window size being the size of the query. Each of the 72 instances was used as the query for a similarity search, and we reported the average results from these experiments. Since the retrieved subsequences may be overlapping, we kept the subsequence with the highest similarity score and dropped any subsequences with an overlapping ratio of >10\%. For each query, we perform top-$k$ search and set $k$ to be the number of ground-truth instances of the task in the dataset, so that precision, recall and F1 scores are equal. We use F1 score and overlap ratio (to the ground-truth subsequences) as performance metrics. 
Table~\ref{tab:rstlocaldist} shows that, compared to the traditional similarity search which retrieves the top-$k$ subsequences with the smallest distance, our proposed procedure clearly improves the retrieval accuracy.  
\begin{table}[ht!]
    \centering
    \begin{tabular}{c|cc}
         Method & F1 Score & Overlap Ratio \\
         \midrule
         Traditional Approach &78.1 & 73.5\% \\
         Proposed Approach & 82.3 & 75.2\% \\
    \end{tabular}
    \caption{Subsequence retrieval accuracy in desktop activity log.}
    \vspace{-1.5em}
    \label{tab:rstlocaldist}
\end{table}

\minihead{Research directions}
We discuss two research challenges for leveraging local distribution. 

\squishitemize
    \item {\em Objective function design.}
    The objective function for traditional top-$k$ search is to maximize $L = \sum_{i=1}^k f_{\text{sim}}(\text{query}, \text{dp}_i)$ where $\text{dp}_i$ denotes one data point. 
    However, when incorporating local distribution, we need an objective function that strikes a balance between instance similarity and local distribution. We provide one example objective function inspired by our proposed approach:
    \begin{equation}
        L'(k) = \sum_{i=1}^k \Bigl(f_{\text{sim}}(\text{query}, \text{dp}_i) + \sum_{j=1}^{i-1} \lambda^{j} f_{\text{sim}}(\text{dp}_j, \text{dp}_i)\Bigr).
    \end{equation}
    Here, $\text{dp}_j$ are examples in the extended query set; $\lambda \in [0, 1]$ acts as a decay factor to reflect the belief that we trust the original query more than the extended query and that we trust the earlier extended queries more than later ones. $\lambda$ therefore provides a knob to balance between instance similarity and local distribution. Our proposed approach can be viewed as a greedy method of optimizing the objective function: we sequentially maximize $L'(1),\dots, L'(k)$, i.e., we keep $\text{dp}_1, \dots, \text{dp}_{i-1}$ fixed when maximizing $L'(i).$.
    \item {\em Improving search efficiency.}
    The proposed search can be resource-intensive as it requires multiple iterations over the dataset for each query, each time with a slightly altered objective. To enhance search efficiency, we could perform the search in small batches, adding multiple near neighbors during each iteration. This strategy would speed up the process, though it might lead to a slight decrease in accuracy. Another way is to adapt existing indexing techniques to consider local distribution signal. For example, we could construct a separate index for density-based clustering. This index could then be cross-referenced with indices designed for classic approximate nearest neighbor searches to factor in both the local density of data points and their proximity to the query during the search. \vspace{-0.5em}
\squishend

\subsection{Leveraging Global Distribution}
\label{sec:global}
In addition to leveraging local distribution near the query to retrieve tightly clustered nearest neighbors, we can also utilize the global distribution. Practitioners have already begun exploring retrieval solutions that consider the global structure of the dataset~\cite{knnvssvm, svm-langchain}.
We discuss a SVM-based example below~\cite{svm-langchain}. 

The main idea is to train a SVM classifier on the entire dataset, wherein the query vector is labeled as the positive example and all other vectors are labeled as negative examples. If multiple positive examples exist, they can be easily incorporated into the training process. 
During the similarity search, data vectors are ranked based on their distances to the separating hyperplane, and the top-$k$ closest one are returned. Vectors on the same side of the hyperplane as the query have negative distances and will be retrieved first. While the traditional similarity search does not consider the global distribution and treats each dimension equally, the SVM classifier aims to identify a hyperplane that effectively separates the positive examples from the negatives. In doing so, the SVM identifies the unique attributes of the positive example within the dataset and uses these distinct features for ranking.

To illustrate, consider Figure~\ref{fig:distri} (2) representing the distribution of people in terms of age (X-axis) and wealth (Y-axis), with the query indicated as the top-right red dot. Traditional similarity search retrieves data points in the query's vicinity (e.g., green dotted circle). In contrast, the SVM-based approach recognizes the query's distinct position in the top right of the global distribution and returns points from the top-right area of the purple curve, suggesting a search for older, wealthy individuals. In applications with high dimensional semantic space, the SVM-based approach can be much more effective in identifying the distinct attributes of the query, thereby facilitating the retrieval of highly relevant and personalized content for the users.

One downside of this approach is the substantial computational cost, as a new classifier needs to be trained on the entire dataset for each query. Potential solutions to mitigate this include:
\squishitemize
    \item {\em Coreset-based solution.} A coreset~\cite{tukan2021coresets} is a small reprensentative sample of the full dataset such that a classifier trained on a coreset mirrors one trained on the full dataset. Therefore, building a coreset reduces the computational effort for training a new classifier for each query. A related research question is how to pre-build the coreset for negative examples when positive examples (the query) are unknown.
    \item {\em Index-based solution.} For some classifiers, it might be possible to build a model parameter index using the negative data points. Specifically, when a new query (positive data point) arrives, learning the parameters of the new model reduces to looking up the parameters in the index using the query.
    \item {\em Symbolic model training.} For some classifiers with closed form solution (e.g., logistic regression), it might be possible to train the model using the negative examples and one symbolic positive example. When the query comes (the values for the positive example is available), we substitute the symbols with the provided values.

\squishend

While we have discussed leveraging local distribution (\S~\ref{sec:local}) and global distribution (\S~\ref{sec:global}) individually, we expect that each is best suited to different applications. Determining the proper usage scenarios for each strategy is also an interesting research question. \vspace{-2em}

\section{Human-in-the-loop Search}
In this section, we discuss opportunities to leverage human insights to improve the accuracy and efficiency of similarity search results. We consider two forms of user feedback: direct labels on the similarity search results (\S~\ref{sec:label}) and user-defined functions that filter similarity search results (\S~\ref{sec:udf}). \vspace{-0.5em}

\subsection{Application Scenarios}
\vspace{-0.5em}
\minihead{Query-by-sketch Video Retrieval} Motion queries are an important class of video analytics queries that focus on the movement patterns and interactions of objects over a sequence of video frames. We are developing a visual query language, \sys, for exploratory motion queries in videos, which allows users to define exploratory motion queries in video analytics by sketching events of interest on a canvas. 
In the backend, these user-drawn sketches are transformed into similarity search queries, identifying pertinent video clips without requiring users to specify low-level details such as time duration, distance threshold, or object relationships.

A key challenge in \sys is the inherent ambiguity in human-drawn sketches. For example, a sketch of a car turning left might leave it unclear if the user is seeking clips where a car initially heads upwards (on screen) before the left turn or simply all left turns, regardless of the car's initial direction.

\begin{figure}[ht!]
    \centering
\includegraphics[width=\linewidth]{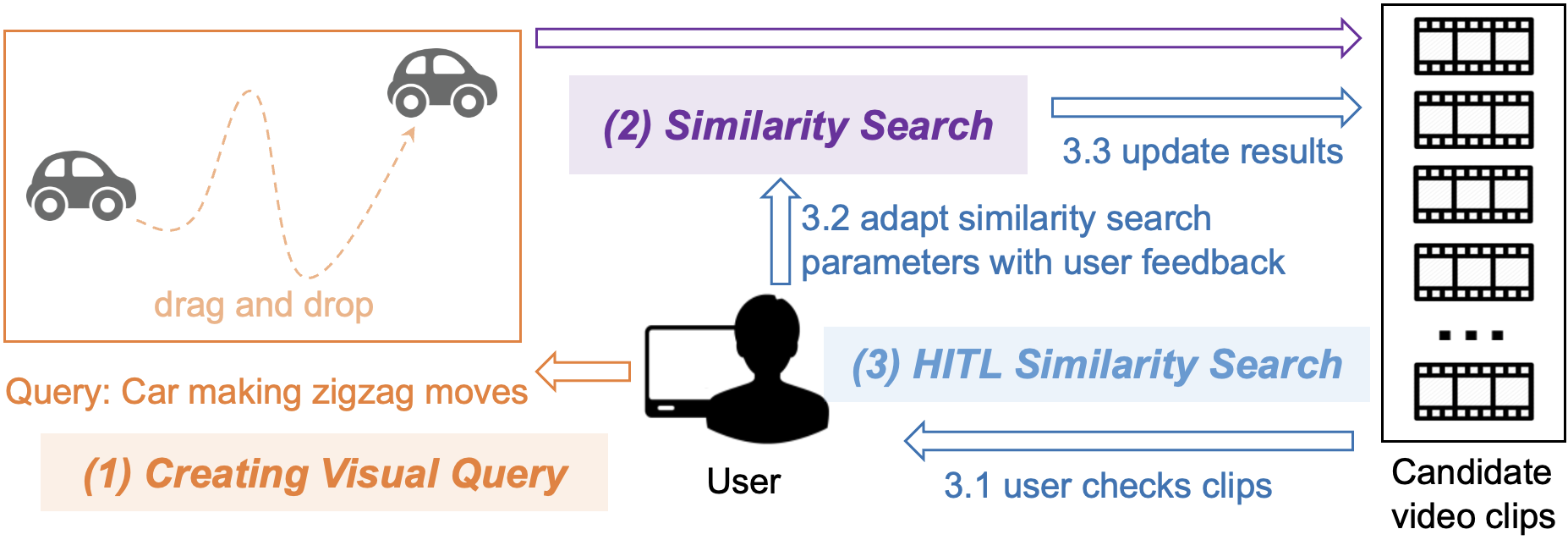}
\caption{Query-by-sketch powered by HITL similarity search.}
\label{fig:HITLSS}
\end{figure}

Therefore, a core component is human-in-the-loop (HITL) similarity search, as shown in Figure~\ref{fig:HITLSS}. Specifically, when the initial similarity search results are presented to the user, they provide feedback on these found examples by labeling them as positive or negative. Based on this feedback, we adapt similarity search parameters (e.g., weights of different embeddings or embedding feature dimensions), and the similarity search results are updated. This process can be repeated multiple times as needed.

\minihead{Data Labeling} Labeling large image or video datasets is a challenging task due to the significant manual effort involved. A promising approach to generate labels efficiently is the use of label propagation based on similarity search~\cite{iscen2019label}. In this workflow, a user labels a few initial examples for a particular class. Subsequently, similarity search retrieves similar examples, which are tentatively labeled as belonging to the same class. Given the potential for false positives, the user reviews the retrieved examples, providing feedback to refine the similarity search and retrieve more accurate matches.

\minihead{Conversational Information Retrieval} Large Language Models (LLMs) facilitate a conversational style of information access. For example, in order to retrieve a specific image from a large collection, a user could engage in a multi-round conversation with the LLM. In each round, the LLM retrieves an image based on the user's description. The user then provides feedback, refining the description if the retrieved image isn't what they were looking for. The LLM then uses this feedback to retrieve a different image. This process naturally forms a HITL similarity search problem. \vspace{-1em}

\subsection{Incorporating User-Provided Labels}
\label{sec:label}
In similarity search, we compare a query $x_{q}$ against a large set of data points ${x_1, \dots, x_n}$. Suppose the similarity search retrieves the top 5 data points, ranked in descending order of similarity: ${x_1, x_2, x_3, x_4, x_5}$. User feedback indicates that ${x_1, x_4, x_5}$ are positive results, and ${x_2, x_3}$ are negative.

The user feedback suggests that $sim(x_{q}, x_4)$ and $sim(x_{q}, x_5)$ should be larger than $sim(x_{q}, x_2)$ and $sim(x_{q}, x_3)$. There are three ways of incorporating this feedback: (1) adapting the embedding of the query; (2) adapting the embedding of the data; (3) adapting the similarity function $sim$. In some cases, it could help to apply a combination or even all methods. We discuss these options below.

\minihead{Adapting query embedding}
This approach is simple and efficient as it requires minimal changes to similarity search. It is also implicitly adopted in conversational information retrieval, as the user's feedback naturally updates the query embedding for subsequent responses. However, it may not always be possible to adjust the query embedding to satisfy user feedback. For example, if embeddings are one-dimensional and $x_2=x_3=2, x_4=1, x_5=3$, with absolute distance as the similarity function, there is no $x_q$ where $sim(x_{q}, x_4)$ and $sim(x_{q}, x_5)$ are larger than $sim(x_{q}, x_2)$ and $sim(x_{q}, x_3)$. Therefore, this approach works best when only minor adaptations are needed. An interesting area for future research would be to formally analyze the scenarios suitable for this method.

\minihead{Adapting data embedding}
This approach is preferable when larger adaptations are required, such as when conducting a series of queries with continuous user feedback. Yet, modifying the embedding of all data points is computationally intensive. We propose two more practical solutions:
\squishitemize
\item {\em Parameterized embedding.} The embedding of each data point $x_i$ is a weighted sum of several high-dimensional embeddings: $x_i = w_1x_1'+\dots + w_mx_m'$. There is no need to instantiate each $x_i$ as we only need store each component $x_i'$ and the weights. These weights, $w_1, \dots, w_m$, can be adjusted to account for user feedback. Existing work in this direction
~\cite{learnglobalweightrecomm2016, weightknn2017} 
could be expanded upon to explore different methods of constructing parameterized embedding.
\item {\em Partial and lazy update.} If we have a million data points and the user provides feedback on just two examples, updating the embedding for all data points is wasteful. Ideally, we want to update only the data points that would affect future query results. One approach is to update only those data points relevant to the query where user feedback is provided (partial update). Another approach is to only materialize the updates when a query arrives for which the updated data embeddings could change top-$k$ search results (lazy update).
\squishend
 
\minihead{Adapting similarity function}
The last option is to have a parameterized similarity function. This is generally not preferred as many methods for accelerating similarity search assume a limited form of the similarity function, such as cosine similarity or L1/L2 distances. Furthermore, the benefits of adapting the similarity function can be largely achieved by adapting the query/data embedding. \vspace{-1em}

\subsection{Incorporating User-Defined Functions}  
\label{sec:udf}
Users can enhance similarity search results by adding direct labels or encoding their expertise and preferences through user-defined functions (UDFs). These UDFs can either filter inputs before the similarity search to improve search efficiency, or filter outputs after the search to enhance retrieval accuracy. This is in line with VIVA's~\cite{viva} approach of optimizing SQL query execution using human insights, though our focus is on similarity search pipelines.

\minihead{Example UDFs} UDFs can come from various sources, as we illustrate in the context of video retrieval applications.
Object detection models can support predicates on objects within a single frame.
For example, a predicate like \lstinline[style=SQLStyle]{ArrayCount(FastRCNNObjectDetector(data).labels, 'cars') > 3} retrieves frames with more than 3 cars. 
Scene-graphs, representing objects within a frame as nodes and relationships between objects as edges, can support more complex filters over a sequence of frames. 
Users might specify that a {\em car} and {\em pedestrian} must be within {\em 10 pixels} for at least {\em 500 frames}. 

Beyond vision models, users can leverage {\em exogenous data sources} as predicates.
For example, in a baseball game, goal-scoring clips can be detected by observing changes in the scoreboard through an OCR model.
Additional data modalities, such as audio, are another source of exogenous data. 
An audio transcript obtained via automatic speech recognition models like Whisper could assist in identifying key moments in a baseball game.

\minihead{Efficient Execution of UDFs} An interesting research question remains on how to use query optimization techniques to generate efficient, accuracy-aware physical plans for similarity search pipelines with UDF-predicates.

Compared to metadata predicates explained in \S~\ref{sec:advance}, UDF-based predicates are generally more computationally expensive. 
For example, object detection model ByteTrack can process around 6 frames per second using 1 GPU. 
Therefore, we can not afford to the materialize UDFs on all input data (e.g., invoking ByteTrack to materialize the label for each video frame) and pre-compute an index. 
Instead, our efforts should be directed towards avoiding unnecessary materialization.

Further, UDFs come with various performance and accuracy trade-offs. For example, scene-graph models, while providing detailed relationship data, generally require more computational resources compared to object detection models. There is also a notable variation among different model architectures for the same task. Hence, the query optimizer can choose from different physical plans based on computational requirements and accuracy goals.

\section{\Multi Similarity Search}
In this section, we introduce a variant of the similarity search problem which we refer to as \multi similarity search. This variant considers the similarity between \emph{two sets of objects}, where objects in each set conform to certain relationship constraints. \vspace{-0.5em}

\subsection{Application Scenarios}
\minihead{Trajectory Matching} Consider the video retrieval application where we wish to identify video clips containing specific object trajectories. Suppose our query clip portrays a soccer player passing a ball to another player, and the video contains bounding box trajectories of four players (\cref{fig:alignment}).
The overall similarity between the query and the video clip is the highest similarity score achieved among all potential alignments of movements between the two players in the query and any two of the four in the video. There are $4\times 3 = 12$ alignments in this example: 4 choices for the player passing the ball, and 3 remaining choices for the receiver.
Traditional methods often require evaluating all possible alignments to determine the highest similarity score between the two clips. As the number of objects in the video or query clip increases, this leads to a significant increase in time complexity.

\minihead{Visual Concept Search} 
Visual concepts are segments of images that carry semantic meaning, such as specific objects (e.g., cars) or components of objects (e.g., a car hood). Each image can contain dozens to hundreds of these concepts, each of which is represented by a unique embedding vector. These vectors can be utilized to locate images containing relevant content~\cite{hoque2022visual, ahn2023escape}. Moreover, a more fine-grained search could involve spatial constraints such as ``below,'' ``next-to,'' or even specific angles and distances. For example, a autonomous-vehicle researcher might want to find scenarios of ``children riding bikes next to cars''. A query for this scenario might look like $((c_{child} \text{ angle[0,30] ON_TOP } c_{bike}) \text{ NEXT_TO } c_{car})$. In this case, the search must identify images that contain three embedding vectors each closely matching $c_{child}, c_{bike}, c_{car}$ respectively, and also verify that the visual concepts in the image align with the specified spatial constraints.

\subsection{Handling Inter-object Constraints}

The \multi search expands upon the multi-vector query problem presented in Milvus~\cite{wang2021milvus}. While both involve multiple vectors in each query, the \multi problem additionally introduces constraints among vectors within the same query.

More formally, consider a query $Q=\{q_1, \dots , q_m\}$ of $m$ ordered objects where each $q_i$ is the feature representation of the $i^{th}$ object and a collection of data points $D=\{\text{dp}_1, \dots, \text{dp}_n\}$. We also have a constraint set $C(Q, R)$ describing the relationship between the objects in the query $Q$ and the result $R$. The goal is to identify $m$ ordered data points $D_{\text{cand}} = \{\text{dp}^{*}_{1}, \dots, \text{dp}^{*}_{m}\}$ that maximizes $sim(Q, D_{\text{cand}})$ while satisfying the constraint set $C(Q, D_{\text{cand}})$. 

\begin{figure}[t]
 \centering  
  \includegraphics[clip, width=0.8\linewidth]{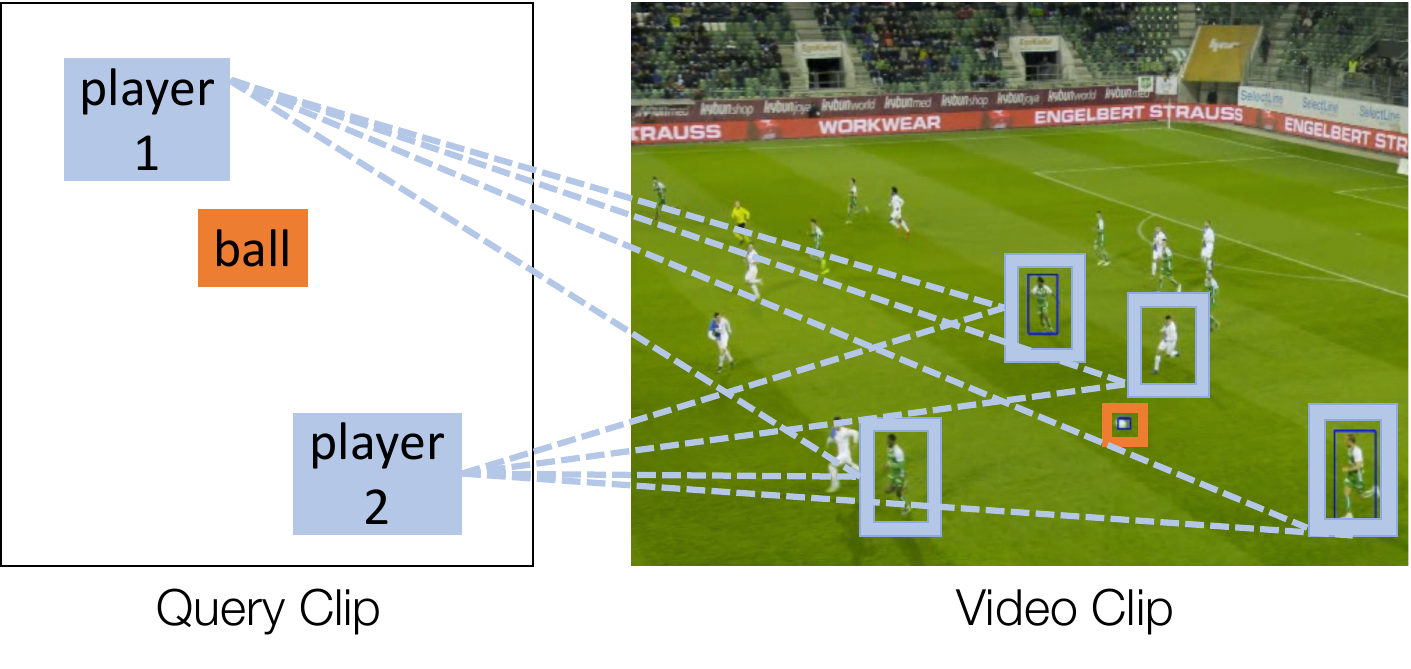}  
    \caption{
      \textbf{There are 12 possible alignments between the player trajectories in the query clip and the video clip.}
    }
  \label{fig:alignment}
\end{figure}

We illustrate this using the example in Figure~\ref{fig:alignment}. Here, the query $Q=\{q_1, q_2, q_3\}$ and the result $D_{\text{cand}} = \{\text{dp}^{*}_{1}, \text{dp}^{*}_{2}, \text{dp}^{*}_{3}\}$ each includes the embedding vectors of the trajectory of two soccer players and the ball. The similarity $sim(Q, D_{\text{cand}})$ is computed as the sum of Euclidean distance between corresponding object embeddings: $sim(Q, D_{\text{cand}})=\sum_{i=1}^3 ||q_i -\text{dp}^{*}_{i}||_2$. The constraint is that all objects in $D_{\text{cand}}$ appear in the same frames and that $q_i$ and $\text{dp}^{*}_{i}$ have the same object class. For instance, $q_3$ and $\text{dp}^{*}_{3}$ are trajectories of the soccer ball, while the remaining vectors denote player trajectories.

\minihead{Research directions} We discuss potential research directions for the \multi similarity search problem. 
\squishitemize
\item {\em Computation Reuse.} In time series applications (e.g., trajectory matching in Figure~\ref{fig:alignment}), when a sliding window is employed for the search, computations from overlapping windows can potentially be reused.
For example, if the top candidates within the time window $[0, 100]$ are object 1 and object 2, it is likely that the same objects continue to be the top candidates upon shifting the window slightly to $[1, 101]$.
Compared to traditional similarity search, the main challenge for reuse is solving object alignment, which involves mapping each candidate object to its corresponding object in the query in an efficient way, while also taking advantage of previous computational results. This known as the assignment problem with changing costs~\cite{mills2007dynamic}, which has polynomial solutions for simple similarity functions, but could be intractable for complex functions.

\item {\em Execution Strategies.}  There are multiple strategies for executing \multi search queries. One approach is to perform similarity search individually for each query object to retrieve top-$k$ candidates $\{\text{dp}_1^i, \dots\}$ (where $k$ can vary); we then join the search results for all queries and enforce the constraint set. Since the individual search can be easily parallelized, the approach is favorable when there are many objects in query and the constraints are simple. For more complex and selective constraints, filtering vectors based on constraints before the similarity search can be more efficient. For instance, in the query for ``children riding bikes next to cars'', we could first identify pairs of concepts that meet the spatial constraint ($(c_{child} \text{ angle[0,30] ON_TOP } c_{bike})$) using a spatial index, then check the candidate vector's distance to the query. A future research question is to develop a query optimizer that can automatically select the most suitable execution strategy.
\squishend

\bibliographystyle{ACM-Reference-Format}
\raggedright
\bibliography{main,vectordb}

\end{document}